1  **Characterization of a novel angular dioxygenase from fluorene-degrading**
2  ***Sphingomonas* sp. strain LB126**

3  Luc Schuler[1], Sinéad M. Ní Chadhain[2], Yves Jouanneau[3], Christine Meyer[3], Gerben J.
4  Zylstra[2], Pascal Hols[4] and Spiros N. Agathos[1*$]

6  [1] Unité de Génie Biologique, Institut des Sciences de la Vie, Université catholique de
7  Louvain, Place Croix du Sud, 2/19, B-1348 Louvain-la-Neuve, Belgium.
8  [2] Biotechnology Center for Agriculture and the Environment, Cook College, Rutgers
9  University, New Brunswick, New Jersey, USA.
10 [3] Laboratoire de Chimie et Biologie des Métaux, iRTSV, CEA, CNRS, Université J. Fourier
11 UMR 5249, CEA-Grenoble, F-38054 Grenoble Cedex 9, France.
12 [4] Unité de Génétique, Institut des Sciences de la Vie, Université catholique de Louvain, Place
13 Croix du Sud 5, B-1348 Louvain-la-Neuve, Belgium.

14 **Running Title:** Angular dioxygenase from *Sphingomonas* sp. LB126

15 **Keywords:** Bioremediation, catabolic gene cluster, polycyclic aromatic hydrocarbons,
16 heterologous expression

17 * Corresponding author.
18 Mailing address:  Unité de Génie Biologique,
19                   Institut des Sciences de la Vie,
20                   Université catholique de Louvain,
21                   Place Croix du Sud, 2/19,
22                   B-1348 Louvain-la-Neuve, Belgium.
23 Phone: +32 10 47 36 44.
24 Fax: 32 10 47 30 62.
25 E-mail: spiros.agathos@uclouvain.be

**ABSTRACT**

In this study, the genes involved in the initial attack on fluorene by *Sphingomonas* sp. LB126 were investigated. The α and β subunits of a dioxygenase complex (FlnA1A2), showing 63% and 51% sequence identity respectively, with the subunits of an angular dioxygenase from Gram-positive *Terrabacter* sp. DBF63, were identified. When overexpressed in *E. coli*, FlnA1A2 was responsible for the angular oxidation of fluorene, fluorenol, fluorenone, dibenzofuran and dibenzo-*p*-dioxin. Moreover, FlnA1A2 was able to oxidize polycyclic aromatic hydrocarbons and heteroaromatics, some of which were not oxidized by the dioxygenase from *Terrabacter* sp. DBF63. Quantification of resulting oxidation products showed that fluorene and phenanthrene were preferred substrates.

# INTRODUCTION

Polycyclic aromatic hydrocarbons (PAHs) are ubiquitous environmental contaminants and are formed during the burning, handling or disposal of organic matter including coal tars, crude oil and petroleum products. There are some natural origins, such as forest fires or natural oil seeps, but PAHs mainly arise from combustion- or oil-related anthropogenic activities. A number of organisms that are able to use PAHs as sole source of carbon and energy have been isolated (6) and bioremediation strategies using these organisms have been proposed (17).

Fluorene, a tricyclic aromatic hydrocarbon containing a five-membered ring, offers a variety of possibilities for biochemical attack. Two of these pathways are initiated by a dioxygenation at the 1,2- (5, 9) or 3,4- positions (5, 10, 27) (Fig. 1). The corresponding *cis*-dihydrodiols undergo dehydrogenation, then *meta*-cleavage. The third route (39, 45) is initiated by monooxygenation at the C-9 position to give 9-fluorenol, which is then dehydrogenated to 9-fluorenone. This route is only productive if a subsequent angular carbon dioxygenation forms 1-hydro-1,1a-dihydroxy-9-fluorenone, leading to phthalate, which is degraded in turn via protocatechuate (11, 27, 45) (Fig. 1).

Sphingomonads have been intensively studied for their ability to degrade a wide range of aromatic hydrocarbons (32, 34, 42, 43, 49, 50). The function and organization of catabolic genes often remain obscure since the genes involved in the degradation of aromatic compounds are not always arranged in discrete operons but are frequently dispersed throughout the genome. *Sphingomonas* sp. LB126 was isolated from PAH contaminated soil and is capable of utilizing fluorene as sole carbon source (3). Fluorene degradation by strain LB126 has been previously investigated (48), but the enzymes that govern the initial attack on fluorene were not identified.

62  Habe et al. (13, 14) showed that the Gram-positive dibenzofuran degrading bacterium
63  *Terrabacter* sp. DBF63, can also oxidize fluorene thanks to a cluster of plasmid-borne
64  catabolic genes. The oxygenase component of an angular dioxygenase complex, encoded by
65  *dbfA1A2*, does not cluster with already known dioxygenases. Few data are available regarding
66  genes involved in fluorene degradation by Gram-negative bacteria. Although many PAH
67  dioxygenases are known to oxidize fluorene, the respective strains could not use fluorene as
68  sole carbon source. Recently, the catabolic plasmid pCAR3 from *Sphingomonas* sp. KA1 was
69  described (41). Genes homologous to *dbfA1A2* were found on pCAR3, as well as all genes
70  necessary for the complete degradation of fluorene, but strain KA1 is unable to grow on
71  fluorene as sole source of carbon. We present here the first report, to our knowledge, of genes
72  governing angular attack on fluorene in Gram-negative bacteria using fluorene as the sole
73  source of carbon and energy.
74

## MATERIALS AND METHODS

**Bacterial strains, plasmids, and media.** *Sphingomonas* sp. LB126, the wild-type strain capable of growing on fluorene as the sole source of carbon and energy (3), was kindly provided by VITO (Vlaamse Instelling voor Technologisch Onderzoek, Belgium). *Escherichia coli* Top10 was used as the recipient strain in all cloning experiments. *E. coli* BL21(DE3) was used for gene expression analysis. PCR amplicons were either cloned into pDrive (Qiagen, Valencia, CA) or pGEM-T-easy vector (Promega, Madison, WI), and pET30f (Novagen, San Diego, CA) was used as expression vector. MM284 minimal medium (26) was used for growing *Sphingomonas* sp. LB126 and was supplemented with phosphate buffer (0.5 M $KH_2PO_4$, 0.5 M $K_2HPO_4$, pH 7.2) instead of Tris buffer. Fluorene was provided as crystals in both Petri dishes and liquid media. LB broth (37) was used as complete medium for growing *E. coli* strains. Solid media contained 2% agar. When needed, ampicillin, streptomycin or kanamycin were added to the medium at 100, 200 and 20 µg/ml, respectively. *Sphingomonas* sp. LB126 was grown at 30°C, and *E. coli* strains were grown at 37°C. Bacterial growth was determined by optical density readings at 600 nm ($OD_{600}$).

**DNA manipulations and molecular techniques.** Total DNA from pure cultures of *Sphingomonas* sp. LB126 was extracted using the Ultra Clean DNA Isolation Kit (MoBio, Carlsbad, CA) following the manufacturer's recommendations or using standard methods (37) when a higher DNA concentration was needed. Plasmid DNA extractions, restriction enzyme digestions, ligations, transformations, sequencing and agarose gel electrophoresis were carried out using standard methods (37).

**Polymerase chain reaction (PCR).** Degenerate primers for amplifying conserved sequences of the gene encoding the angular dioxygenase were used as described elsewhere (15). PCR products were purified and cloned into either the pGEM-T or pDrive plasmids. The RT-PCR

101  reactions were performed in 25 μl with 5 ng of total RNA and 20 pmol of each primer with
102  OneStep RT-PCR Kit (Qiagen, Belgium). Total RNA extractions were performed using the
103  RNeasy kit (Qiagen, Valencia, CA) and further purified by spin column and DNase I
104  treatment according to the manufacturer's instructions. The thermocycler program used for
105  the RT-PCR reactions was as follows: 60°C for 30 min, 94°C for 15 min, 30 cycles (94°C for
106  30 s, 50°C for 30 s, 72°C for 45 s), and 72°C for 7 min.
107
108  **Southern Blot detection of catabolic genes**. Genomic DNA (2 μg) was digested with either
109  BamHI, NotI, NsiI or a combination of these enzymes, separated by gel electrophoresis, then
110  blotted onto a positively charged nylon membrane (Amersham, Buckinghamshire, UK) using
111  standard protocols (37). For Southern Blot detection a PCR-amplified DIG-labeled probe was
112  prepared according to the manufacturer's recommendations (Roche Diagnostics, Mannheim,
113  Germany). Pre-hybridization and hybridization were carried out at 68°C. After hybridization,
114  the membrane was washed twice with 2 x SSC (20 x SSC: NaCl, 3 mol/l; Na-citrate, 0.3
115  mol/l; pH 7.0) containing 0.1% sodium dodecyl sulfate (SDS) (w/v) for 5 min at room
116  temperature and twice with 0.1 x SSC containing 0.1% SDS for 15 min at 68°C. Detection
117  was carried out following standard protocols (37). To isolate catabolic genes, total DNA (10
118  μg) was digested with BamHI and NsiI separated by gel electrophoresis and DNA fragments
119  of about 7 kb recovered from the agarose gel. The obtained DNA was cloned into pGEM5Z
120  (Promega) and transformed into *E. coli* Top10. Resulting clones were screened by PCR using
121  the above-mentioned primers.
122
123  **Construction of plasmids for protein overexpression.** Construction of the plasmids used in
124  this study involved multiple PCR amplifications and cloning steps. The *flnA1A2* fragment
125  (1842 bp) was amplified by PCR with the primer pairs: 5'-
126  *CATATG*GCCACAGCCCTCATGAACCACCC-3' and 5'-

*AAGCTT*GGCGCTCACAGGAACACCG-3', introducing NdeI and HindIII sites (italics) at the ends of the amplicon. The PCR product was cloned into pDrive (Qiagen), sequenced, then subcloned into the NdeI and HindIII sites of expression vector pET-30f (Novagen). This construct was transformed into *E. coli* BL21(DE3) for expression analysis.

**Sodium dodecyl sulfate-polyacrylamide gel electrophoresis (SDS-PAGE).** Bacterial cells were pelleted by centrifugation and washed with 10 ml ice-cold phosphate buffer (140 mM NaCl, 10 mM $Na_2HPO_4$, 2.7 mM KCl, 1.8 mM $NaH_2PO_4$, pH 7.4). To the pellet was added 1 ml of ice-cold phosphate buffer and 550 μl of the suspension was subjected to sonication on ice for 20 s (5 s pulse interval; 40% of maximum amplitude). After centrifugation the supernatant and the pellet were mixed with an equal volume of loading solution. SDS-PAGE was performed on mini gels containing 13.3 % polyacrylamide. After electrophoresis, protein staining was performed with Coomassie brilliant blue R-250.

**Dioxygenase overexpression and in vivo assays.** Strain BL21(DE3)(pET30f*flnA1A2*) was grown overnight in 3 ml LB medium with the suitable antibiotics. This culture was used to inoculate 25 ml LB medium (0.1% by volume), which was incubated at 42°C until an $OD_{600}$ of 0.5. IPTG was added to a final concentration of 0.5 mM. The cells were further incubated for 7 h at 25°C. For in vivo assays, cells were centrifuged, washed and resuspended to an $OD_{600}$ of approximately 2 in M9 (37) medium containing 0.2% glucose. Cells (25 ml) were incubated for 48 h at 25°C with 5 ml silicone oil containing 0.1 g/l of each tested PAH.

**GC-MS analysis of PAH oxidation products.** Water-soluble products resulting from PAH oxidation were extracted from the aqueous phase of bacterial suspension by using, columns filled with reverse phase-bonded silica (Upti-clean C18U, 0.5 g, Interchim, Montluçon, France). Columns were washed with 10 ml water then eluted with 1 ml ethyl acetate. The solvent was dried over sodium sulfate and evaporated under nitrogen gas. The dried extracts were then dissolved in 100 or 200 μl acetonitrile, before being derivatized with *N,O-bis*(trimethylsilyl)trifluoroacetamide:trimethylchlorosilane (BSTFA) or *n*-butylboronate (NBB). In order to quantify the dihydrodiols formed upon incubation of BL21(DE3)(pET30f*flnA1A2* recombinant cells with PAHs, 2.3-dihydrobiphenyl (Sigma-Aldrich) was added to 0.1 μM final concentration in the aqueous phase prior to solid phase extraction, and was used as an internal standard. After derivatization and GC-MS analysis, NBB dihydrodiol derivates were quantified on the basis of peak area using a calibration curve generated by analyzing known amounts of anthracene 1,2-dihydrodiol. GC-MS analysis of trimethylsilyl derivatives was carried out as previously described (18). NBB derivatives were separated on MDN-12 capillary column (30 m, 0.25 mm internal diameter; Supelco) using helium as carrier gas at 1 ml/min. The oven temperature was held at 75°C for 1 min, then increased to 300°C at 14°/min, and held at 300° for 8 min. The mass spectrometer was operated in the selected ion monitoring mode, selecting *m/z* values corresponding to the expected masses ($M^+$) of the dihydrodiol derivatives (228 for naphthalene, 278 for anthracene and phenanthrene). The NBB derivative of trihydroxybiphenyl, the oxidation product of dibenzofuran, was monitored at a *m/z* value of 268. The fluorene derivative was detected at a *m/z* of 196 ($M^+$- $OBC_4H_9$), because in contrast to other dihydrodiol derivatives, the abundance of the $M^+$ ion was very low.

**DNA and protein sequence analysis.** Sequence analysis was performed using the DNASTAR software package (Lasergene Inc., Madison, WI*)*. The BLAST search tool was used for homology searches (1). Multiple alignments and phylogenetic trees were produced using the DNASTAR and MEGA3.1 softwares (23).

178 **Nucleotide sequence accession number.** The nucleotide sequence described in this report
179 has been deposited in the Genbank database under accession number EU024110.
180

        **RESULTS AND DISCUSSION**

**Cloning and sequence analysis of genes encoding a novel angular dioxygenase.**

*Sphingomonas* strain LB126 has been studied for its ability to grow on fluorene and degrade phenanthrene, anthracene and fluoranthene by cometabolism (47). In order to detect genes potentially involved in the initial attack of PAHs, a PCR strategy was chosen. The genes involved in fluorene oxidation in strain LB126 were expected to display some similarity with counterparts already described in other PAH degrading *Sphingomonas* strains. Many primer pairs corresponding to conserved domains of previously described PAH dioxygenases were tested (7, 19, 24, 28), but no amplification could be obtained (data not shown). Given the dearth of information regarding fluorene degradation genes in Gram-negative bacteria, primers specific to angular dioxygenase genes from Gram-positive bacteria were tested. Using a set of such primers (15) and total DNA as a template, a 267 bp DNA fragment was amplified, which upon sequencing and translation, revealed 57 % protein sequence identity with a peptide internal to the dibenzofuran 4,4a-dioxygenase α subunit of *Terrabacter* sp. DBF63 (20). The 267 bp fragment was then used as a DIG labeled probe in Southern blot experiments on whole genome extracts of strain LB126. A 6.9 kb fragment encoding four entire open reading frames (ORF) (ORFs 3-6) and three truncated ones (ORFs 1,2 and 7) was recovered (Table 1). ORF1 did not share amino acid sequence similarities with any previously described fluorene catabolic genes, but showed significant homology to TonB-dependent receptor CirA from *Sphingomonas wittichii* strain RW1 (36%) and *Novosphingobium aromaticivorans* F199 (34%). ORF2 encoded a truncated transposase, suggesting that the adjacent gene cluster was probably acquired by horizontal transfer although no change in GC-content was noticed. ORFs 3-7 showed a genetic organization similar to that of the dibenzofuran catabolic operon from *Terrabacter* sp. DBF63 (20) (Fig. 2). Nevertheless, the product of ORF3, a putative dehydrogenase, did not share significant protein sequence similarity with its counterpart (FlnB) from strain DBF63. The highest degree of similarity was found with putative dehydrogenases identified in whole genome sequencing projects of *Mycobacterium* strains MCS and KMS. ORF4 and ORF5 encode the α and β subunits of a putative angular dioxygenase. Their amino acid sequence showed moderate identity (63% and 51%) with DbfA1 and DbfA2 from strain DBF63. Phylogenetic analysis revealed that the ORF4 product did not cluster with dioxygenase α subunits from other sphingomonads, and was only distantly related to the angular dioxygenase from *Sphingomonas witichii* strain RW1 (4). The closest homologues within Sphingomonads were the dioxygenase α subunits from the carbazole-degrading strains *Sphingomonas* sp. KA1 (36 % of protein identity) (41) and *Sphingomonas* sp. CB3 (35 % of protein identity) (40). Interestingly, strain KA1 (41) harbors genes whose products were predicted to catalyze protocatechuate degradation, similar to the *lig* genes of *Sphingomonas paucimobilis* SYK-6 (25) and the *fld* genes of *Sphingomonas* sp. LB126 (48). It appears the genes involved in the initial oxidation of fluorene are more closely related to genes from Gram-positive bacteria and that the genes involved in the degradation of protocatechuate are more conserved in *Sphingomonas* species. ORF5 shows moderate protein identity (48%) to the β subunit DbfA2 from Gram-positive dibenzofuran-degrading *Paenibacillus* sp. YK5 (16). DbfA1 and DbfA2 from strain YK5 are the two subunits of a dioxygenase able to oxidize dibenzo-*p*-dioxin, dibenzothiophene, fluorene, and fluoren-9-one, compounds that could however not be utilized as growth substrates by strain YK5 (16). Transcriptional expression of ORF4 was studied by RT-PCR. Total RNA was extracted from cultures of *Sphingomonas* sp. LB126 grown in the presence of glucose or fluorene. The primers used previously to amplify a 267-bp internal fragment of ORF4 were employed to detect the same portion of cDNA. Results indicated that ORF4 expression was manifold up-regulated in the presence of fluorene (data not shown).. Based on this finding and the observation that ORF4 and ORF5 are the subunits of an angular dioxygenase component that preferentially use fluorene as substrate (see below), we suggest that the two ORFs are involved in the initial attack on fluorene. They were called *flnA1* and *flnA2*. The proteins encoded by *flnA1A2* from *Sphingomonas* sp. LB126 were quite unique, since no functional counterpart had been described so far in Gram-negative bacteria. ORF6, located downstream of *flnA1A2*, showed 42% identity with FlnE, a *meta*-cleavage product hydrolase from strain DBF63 (14). The truncated ORF7 showed similarity to a counterpart from strain DBF63 (FlnD1), encoding an extradiol dioxygenase α subunit. Since *flnD1* from strain LB126 lacks a 3' region, no conclusive homology search could be carried out. Altogether, our findings indicate that the catabolic gene cluster present in strain LB126 might have been inherited by lateral transfer from other genera of PAH-degrading bacteria (Fig. 2) (33).

**Functional expression of FlnA1A2 in *E. coli*.**

In order to study the catalytic activity of FlnA1A2, the corresponding genes were introduced into pET30f and expressed in *E. coli* BL21(DE3). Protein extracts from IPTG-induced cells were separated by SDS-PAGE. The cells overproduced two polypeptides with $M_r$ of 45,000 and 14,000, that did not match exactly the expected sizes of FlnA1 and FlnA2 as calculated from the deduced polypeptide sequence (49.5 and 19.4 kDa). Differences between the theoretical and apparent molecular masses upon SDS-PAGE gels were also observed for the DbfA1 and DbfA2 dioxygenase components from strain DBF63 (20). Significantly, it was found that the recombinant proteins were inactive and mostly insoluble (Fig. 3). When the recombinant strain was grown at 42°C up to an $OD_{600}$ of 0.5, then subjected to IPTG induction at room temperature, a greater proportion of the FlnA1 and FlnA2 proteins was recovered in the soluble fraction (Fig. 3). In order to assess the catalytic activity of FlnA1A2 in *E. coli*, biotransformation assays were carried out using induced cells incubated separately with fluorene, carbazole, dibenzofuran, dibenzothiophene and dibenzo-*p*-dioxin, as well as with representative PAHs. Water-soluble oxidation products released into the culture medium were extracted and analyzed using GC-MS. The detection of PAH oxidation products demonstrated that the recombinant enzyme was active in vivo (Table 2), suggesting that it recruited unspecific electron carriers from the host for function. When strain BL21(DE3)(pET30f), which lacked FlnA1A2, was incubated with the same PAHs under identical conditions, no oxidation product could be detected, demonstrating that FlnA1A2 was responsible for PAH transformation (Table 2).

**Substrate range of FlnA1A2.**

The substrate range of FlnA1A2 was investigated and compared with those of the well-studied angular dioxygenases DFDO (dibenzofuran 4,4a-dioxygenase) from *Terrabacter* sp. strain DBF63 (20) and CARDO (carbazole 1,9a-dioxygenase from *Pseudomonas resinovorans* sp. CA10 (31, 38). When fluorene was used as substrate, three oxidation products could be detected (Table 2). This could be due to the limited activity of FlnA1A2 since no specific ferredoxin nor ferredoxin reductase were expressed at the same time. The major product was identified as 1-hydro-1,1a-dihydroxy-9-fluorenone (63 %) based on the *m/z* fragment pattern of its mass spectrum, which was identical to that of the DFDO-mediated oxidation product of fluorene (20, 27) and 9-fluorenol by CARDO (44). Moreover the conversion ration of 9-fluorenol by CARDO was lower in comparison with dibenzofuran and carbazole. Interestingly CARDO does not yield 1-hydro-1,1a-dihydroxy-9-fluorenone when fluorene is used as substrate (44). The oxygenation for fluorene including monooxygenation and lateral dioxygenation was hard to be catalyzed by CARDO suggesting that fluorene is not a preferable substrate for CARDO (44). Fluorenol-dihydrodiol (7 %) and dihydroxyfluorene (29 %) were also produced by FlnA1A2 from strain LB126. The latter product was not formed by DFDO. Fluorenol-dihydrodiol probably resulted from spontaneous transformation of 1-hydro-1,1a-dihydroxy-9-fluorenone since this product was not detected after short incubations. Fluorenol is likely oxidized to fluorenone by a non specific dehydrogenase from *E. coli*. Indeed, we also observed such a spontaneous oxidation upon incubation of fluorenol with the control strain BL21(DE3)(pET30f) lacking the *flnA1A2* construct. Therefore, a dehydrogenase is probably not essential to transform fluorenol to fluorenone but may be required *in vivo* to catalyze the reaction at a reasonable rate. 1-Hydro-1,1a-dihydroxy-9-fluorenone also accumulated when fluorenol or fluorenone were used as substrates, showing that FlnA1A2 was involved in at least two steps in fluorene catabolism (Fig. 1). Since no specific ferredoxin or ferredoxin reductase was expressed at the same time no fluorenol or fluorenone was detected. Given the low activity of FlnA1A2 and the necessity of monooxygenation before angular dioxygenation can occur, fluorenol and fluorenone are probably instantly consumed and are therefore not present.

Three heteroatomic analogs of fluorene, i.e. dibenzofuran, carbazole and dibenzothiophene were tested as substrates for angular oxidation. Dibenzofuran was transformed into 2,2',3-trihydroxybiphenyl by FlnA1A2, as previously found for DFDO (20) and CARDO (31). The initial attack occurred at the 4 and 4a carbon atoms as put forward by Fortnagel et al. in 1989 (8). The dioxygenation of dibenzofuran produces a highly unstable hemiacetal product that could not be observed. Incubation with dibenzothiophene produced traces of dibenzothiophene-sulfoxide and dibenzothiophene-sulfone. These metabolites were previously identified as metabolic intermediates of dibenzothiophene degradation by *Brevibacterium sp.* DO (46), DFDO (27) and CARDO (31). Since FlnA1A2 was able to perform angular dioxygenation on fluorene and dibenzofuran, hydroxylation of dibenzothiophene-sulfone at the angular position was expected. The activity of the enzyme towards dibenzothiophene might have been too low to detect an angular dioxygenation product by GC-MS. Even though carbazole is a structural analogue of fluorene, no angular oxidation product could be identified. The crystal structure of CARDO bound with carbazole was solved and a molecular mechanism of angular dioxygenation for carbazole was proposed (2). Given the low protein identity between CARDO and FlnA1 (16%) no hypothesis could be established why FlnA1A2 does not perform angular dioxygenation on carbazole. Mono- and dihydroxycarbazole were the only oxidation products detected by GC-MS. DFDO from *Terrabacter* sp. DBF63 was not able to perform angular dioxygenation on this substrate. Detection of monohydroxycarbazole suggests that FlnA1A2 transforms carbazole to the corresponding dihydrodiol by lateral dioxygenation. Resnick et al. reported that carbazole dihydrodiols are unstable and spontaneously form monohydroxycarbazole by dehydration (35). CARDO released 2'-aminobiphenyl-2,3-diol upon angular oxidation of carbazole (31). Incubation with dibenzo-*p*-dioxin yielded 2,3,2'-trihydroxydiphenylether via angular dioxygenation based on the *m/z* fragments described from DFDO and CARDO.

Since *Sphingomonas* sp. LB126 is able to use phenanthrene, fluoranthene and anthracene in cometabolic degradation (47), we tested whether FlnA1A2 would attack these PAHs. *cis*-9,10-Dihydroxy-9,10-dihydrophenanthrene, previously identified as a product formed by pyrene dioxygenase from *Mycobacterium* 6PY1 (22), was detected as the major oxidation product of phenanthrene. Interestingly, *cis*-3,4-dihydroxy-3,4-dihydrophenanthrene which is produced in the catabolic pathway of known phenanthrene degraders including sphingomonads (7, 34, 50) was not formed. Monohydroxyphenanthrene was detected in low amounts (4 %) and might have resulted from spontaneous dehydration of the corresponding dihydrodiol. In contrast, DFDO did not produce any metabolite when incubated in the presence of phenanthrene (21). When incubated with fluoranthene, trace amounts of monohydroxyfluoranthene could be detected. Anthracene yielded three metabolites. The major compound could be identified as *cis*-1,2-dihydroxy-1,2-dihydroanthracene by comparison to the oxidation product formed by Phn1 from *Sphingomonas* sp. CHY-1 (7). Trace amounts of monohydroxyanthracene were also present. CARDO produced the same metabolites but DFDO did not. Moreover, a second putative anthracene-diol could be identified. Its mass spectrum was similar to that of *cis*-1,2-dihydroxy-12-dihydroanthracene but the retention time was different. Since no angular attack on anthracene is possible without a preliminary monooxygenation, we suggest that this compound could be *cis*-2,3-dihydroxy-2,3-dihydroanthracene. This metabolite has not been produced by any other enzyme reported so far. When incubated with biphenyl or naphthalene, FlnA1A2 produced the well known metabolites also reported for DFDO and CARDO (20, 31). Our results show that FlnA1A2 from strain LB126 is unique in that it shares characteristics with both DFDO and CARDO.

The catalytic activity of FlnA1A2 towards fluorene and other PAHs was compared by estimating the amount of di- or trihydroxylated products formed by strain BL21(DE3)(pET30f*flnA1A2* after overnight incubation. Products were extracted and quantified as NBB derivatives by GC-MS analysis as described in Materials and Methods. Results showed that 1-hydro-1,1a-dihydroxy-9-fluorenone (97.5 μM) and 9,10-phenanthrene dihydrodiol (96.3 μM) accumulated at highest concentrations indicating that fluorene and phenanthrene were preferred substrates (Table 3). GC-MS data on NBB derivatives confirmed that FlnA1A2 attacked fluorene in angular position and generated 9,10-phenanthrene dihydrodiol instead of the more common 3,4-isomer. In this respect, the activity of FlnA1A2 is quite different from that of other known phenanthrene dioxygenases. In addition, FlnA1A2 showed a relatively low activity with naphthalene. Dibenzofuran and dibenzo-*p*-dioxin apparently yielded low amounts of products, essentially because the trihydroxylated compounds generated from these substrates reacted poorly with NBB (data not shown). The amount of the trihydroxylated products was therefore tentatively determined based on the peak area of the trimethylsilyl derivatives using 2,3-dihydroxybiphenyl as standard (Table 3). These results, together with the fact that neither phenanthrene nor dibenzofuran can support growth of strain LB126, provide additional evidence that FlnA1A2 acts as an angular dioxygenase specifically dedicated to fluorene initial attack.

The initial step in the aerobic bacterial degradation of PAHs is the introduction of two hydroxyl groups into the benzene ring, forming *cis*-dihydrodiols. Dioxygenases usually perform oxygenation at lateral positions. This has been described in detail for naphthalene and phenanthrene (6). Some information is available regarding initial dioxygenases from sphingomonads, such as those encoded by the *bphA1A2f* genes from *Novosphingobium* *aromaticivorans* sp. F199 (36), *Sphingobium yanoikuyae* B1 (29), and the *phnA1aA2a* genes from *Sphingomonas* sp. CHY-1 (18). These strains are able to oxidize fluorene but cannot use it as sole source of carbon and energy. The enzymes involved in fluorene oxidation in strain LB126 show relatively high degrees of sequence identity with proteins from Gram-positive bacteria, and were likely acquired by lateral gene transfer since a truncated transposase was identified upstream of the catabolic genes. In angular dioxygenation the carbon atom bonded to the carbonyl group in 9-fluorenol and the adjacent carbon atom in the aromatic ring are both oxidized. FlnA1A2 was able to perform monooxygenations in which the methylene carbon atom in fluorene or the sulfur atom in dibenzothiophene were oxidized. This is an essential step to increase the electron withdrawing capabilities necessary for angular dioxygenation to occur. In dibenzofuran, dibenzo-*p*-dioxin and carbazole, the connecting atoms, O and N respectively, have high electronegativities and these compounds must not be oxidized before angular dioxygenation (30). FlnA1A2 was most active in the presence of fluorene and dibenzofuran. The limited activity towards other PAHs could explain the necessity for a second carbon source to support growth. FlnA1A2 from *Sphingomonas* sp. LB126 was able to perform monooxygenations, angular and lateral oxygenations on PAHs and heteroaromatics that were not oxidized by DFDO from *Terrabacter* sp. DBF63.

## ACKNOWLEDGEMENTS

L.S. gratefully acknowledges the Fund for the Promotion of Research in Industry and Agriculture (F.R.I.A.), Belgium, for providing a doctoral fellowship. L.S. also wishes to thank the members of the Unit of Physiological Biochemistry (FYSA), Catholic University of Louvain, for their daily help and constructive remarks for many years.

**Figure Legends**

**Figure 1.** Proposed pathways for fluorene degradation and bacteria involved. 1, *Arthrobacter* sp. strain F101 (5, 9); 2, *Terrabacter* sp. strain DBF63 (27); 3, *Brevibacterium* sp. strain DPO1361 (45); 4, *Pseudomonas* sp. strain F274 (11); 5, *Burkholderia cepacia* F297 (12); 6, *Sphingomonas* sp. strain LB126 (48).

**Figure 2.** Genetic organization of the 6.9 kb DNA region containing fluorene catabolic genes in *Sphingomonas* sp. LB126 compared to *Paenibacillus* sp. YK5 (AB201843), *Terrabacter* sp.YK3 (AB075242), *Rhodococcus* sp. YK2 (AB070456) and *Sphingomonas* sp. KA1 (NC_008308) and *Terrabacter* sp. DBF63 (AP008980). The arrows indicate location and direction of transcription of the ORFs. Black arrows represent genes involved in the initial attack on fluorene; dark gray arrows indicate genes involved in the electron transport chain or phthalate degradation (*pht*), white arrows indicate regulatory genes and light gray arrows represent genes not directly involved in fluorene oxidation. ' Truncated ORF

**Figure 3.** Detection of FlnA1 and FlnA2 overproduced in *E. coli* BL21(DE3). Soluble (supernatant) and insoluble proteins (pellet) were analysed on SDS-PAGE. *E coli* harbouring pET30f lacking the *flnA1A2* insert (V) was used as control. Protein extracts from cells, overexpressing FlnA1A2 (V+I), grown at 37°C and 42°C up to an OD$_{600}$ of 0.5 prior to IPTG induction are shown. Arrows indicate the α and β subunits of the angular dioxygenase. Molecular mass (kDa): New England Biolabs, Prestained Protein Marker, Broad Range.

569

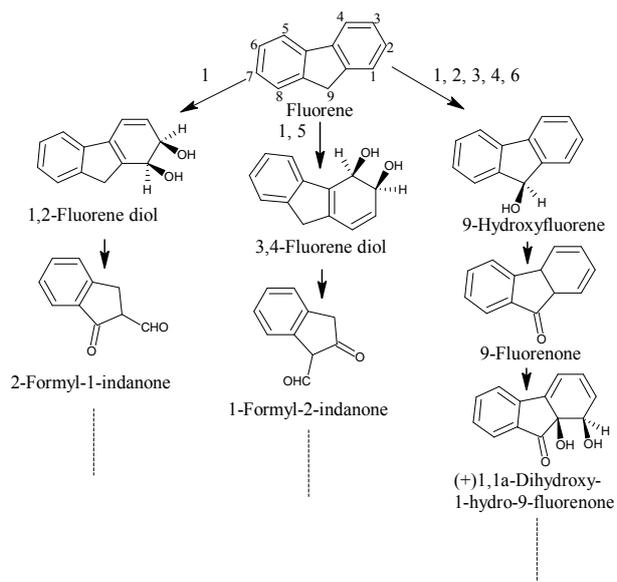

570

571 Fig. 1

572

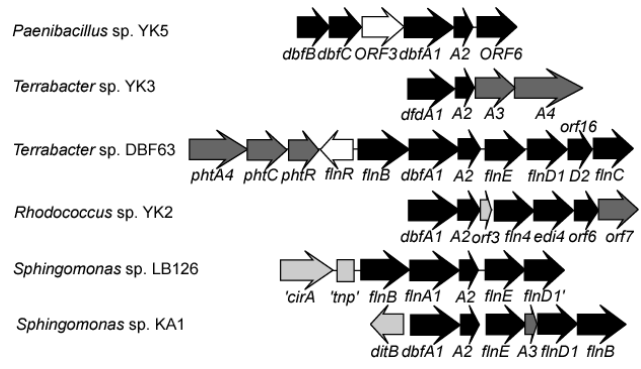

573

574 Fig. 2

575

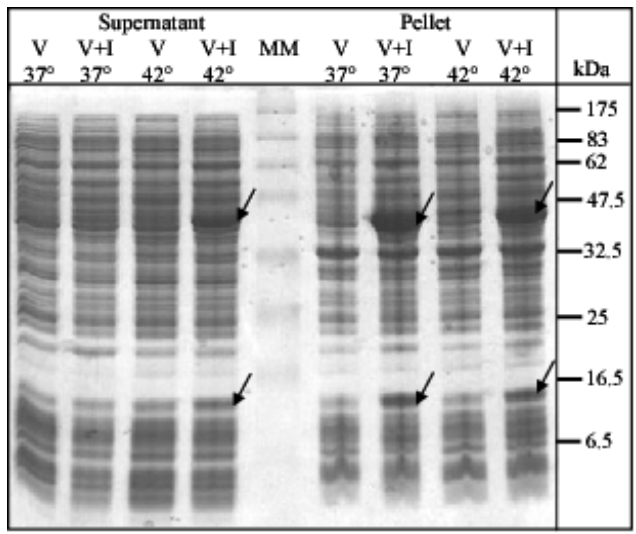

576

577 Fig. 3

Table 1. Homology search analyses of the recovered ORFs from fluorene-degrading *Sphingomonas* sp. strain LB126.

| ORF | Gene | Probable function or product | Homologous protein | Source | Identity (%) | Accession number |
|---|---|---|---|---|---|---|
| ORF1 | '*cirA* | TonB-dependent receptor | CirA | *Sphingomonas wittichii* RW1 | 36 | YP_001262040 |
| | | | CirA | *Novosphingobium aromaticivorans* DSM 12444 | 34 | YP_001165948 |
| ORF2 | '*tnp*' | Transposase | Transposase | *Mesorhizobium loti* MAFF303099 | 60 | NP_085624 |
| | | | Transposase | *Sinorhizobium medicae* WSM419 | 57 | EAU08642 |
| ORF3 | *flnB* | Probable dehydrogenase | probable dehydrogenase | *Mycobacterium* sp. MCS | 40 | ABG07792 |
| | | | probable dehydrogenase | *Mycobacterium* sp. KMS | 40 | ZP_01286209 |
| | | | probable dehydrogenase | *Rhodobacterales* sp. HTCC2654 | 28 | ZP_01014534 |
| ORF4 | *flnA1* | angular dioxygenase α subunit | DbfA1 | *Terrabacter* sp. DBF63 | 63 | BAC75993 |
| | | | DbfA1_YK2 | *Rhodococcus* sp. YK2 | 54 | BAC00802 |
| | | | DbfA1 | *Paenibacillus* sp. YK5 | 52 | BAE53401 |
| ORF5 | *flnA2* | angular dioxygenase β subunit | DbfA2 | *Rhodococcus* sp. YK2 | 52 | BAC00803 |
| | | | DbfA1YK2 | *Terrabacter* sp. DBF63 | 51 | BAC75994 |
| | | | DbfA2 | *Paenibacillus* sp. YK5 | 48 | BAE53402 |
| ORF6 | *flnE* | hydrolase | FlnE | *Terrabacter* sp. DBF63 | 42 | BAE45094 |
| | | | ORF4 | *Rhodococcus* sp. YK2 | 42 | BAC00805 |
| | | | A/b hydrolase_1 | *Mycobacterium* sp. MCS | 30 | YP_642596 |
| ORF7 | *flnD1*' | extradiol dioxygenase α subunit | FlnD1 | *Terrabacter* sp. DBF63 | 12 | BAC75996 |
| | | | BphC6 | *Rhodococcus rhodochrous* | 12 | BAD10908 |
| | | | Edi4 | *Rhodococcus* sp. YK2 | 12 | BAC00806 |

579 Table 2. PAH selectivity of FlnA1A2 from *Sphingomonas* sp. LB126 as expressed in *E. coli* and comparison to DFDO (20) and CARDO (31).

| Substrate | Possible products | Principal fragment ions[a] | Retention Time (min) | Yield (%)[b] | DFDO[c] | CARDO[c] |
|---|---|---|---|---|---|---|
| Naphthalene | *cis*-1,2-Dihydroxy-1,2-dihydronaphthalene[d] | 306 (8), 275 (5), 203 (32), 191 (100) | 13.455 | 92.9 | + | + |
| | 1-naphthol[d,g] | 216 (86), 201 (100), 185 (46), 141 (24) | 12.375 | 7.1 | + | + |
| Biphenyl | *cis*-2,3-Dihydroxy-2,3-dihydrobiphenyl[d] | 332 (52), 243 (22), 227 (100), 211(18) | 15.134 | 83.9 | + | + |
| | 2-Hydroxybiphenyl[d,g] | 242 (48), 227 (76), 211 (100), 165 (7), 152 (20) | 12.910 | 8.5 | + | + |
| | 3-Hydroxybiphenyl[d,g] | 242 (74), 227 (100), 211 (47), 165 (8), 152 (22) | 14.214 | 7.6 | + | + |
| Phenanthrene | *cis*-9,10-Dihydroxy-9,10-dihydrophenanthrene[e] | 356 (16), 253, 191, 147 (100), 73 (99) | 16.728 | 95.1 | - | + |
| | Monohydroxyphenanthrene[g] | 266 (100), 251 (65), 235 (27), 176 13) | 17.464 | 4.1 | - | + |
| Anthracene | *cis*-1,2-Dihydroxy-1,2-dihydroanthracene[d,f] | 356 (5), 266 (13), 253 (34), 191 (62), 147 (26), 73 (100) | 17.348 | 68.5 | - | + |
| | Anthracenedihydrodiol | 356 (34), 266 (82), 253 (3), 191 (3), 147 (60), 73 (100) | 17.874 | 26.8 | - | - |
| | Monohydroxyanthracene[g] | 266 (79), 251 (14), 235 (6), 191 (6), 165 (12), 73 (100) | 17.260 | 4.7 | - | + |
| Fluorene | Dihydroxyfluorene | 342 (14), 253 (46), 152 (17), 73 (100) | 18.477 | 28.9 | - | + |
| | 1-Hydro-1,1a dihydroxy-9-fluorenone[d] | 358 (65), 253 (59), 147 (36), 73 (100) | 16.788 | 63.6 | + | - |
| | Fluorenol-dihydrodiol | 360 (39), 270 (95), 242 (100), 181 (55), 165 (13) | 16.455 | 7.4 | - | - |
| 9-Fluorenol | 1-Hydro-1,1a-dihydroxy-9-fluorenone[d] | 358 (40), 253 (41), 147 (39), 73 (100) | 16.800 | 82 | + | + |
| | Fluorenol-dihydrodiol | 360 (34), 270 (90), 242 (100), 181 (78), 165 (23) | 16.459 | 18 | - | - |
| 9-Fluorenone | 1-Hydro-1,1a-dihydroxy-9-fluorenone[d] | 358 (56), 253 (51), 147 (43), 73 (100) | 16.795 | 76 | + | + |
| | Fluorenol-dihydrodiol | 360 (30), 270 (91), 242 (100), 181 (85), 165 (24) | 16.459 | 24 | - | - |
| Fluoranthene | Monohydroxyfluoranthene[g] | 290 (81), 275 (54), 259 (47), 215 (76) | 20.038 | 100 | Not tested | + |
| Carbazole | Monohydroxycarbazole[g] | 255 (100), 239 (51), 224 (47), 209 (22), 166 (11) | 17.128 | 56.4 | - | - |
| | Dihydroxycarbazole | 343 (100), 327 (34), 252 (7), 164 (2) | 18.688 | 9.7 | - | - |
| | Monohydroxycarbazole[g] | 327 (100), 312 (24), 165 (1), 73 (39) | 18.997 | 33.9 | - | - |
| Dibenzofuran | 2,2',3-Trihydroxybiphenyl[d] | 418 (50), 403 (5), 315 (70), 73 (100) | 16.357 | 100 | + | + |
| Dibenzo-*p*-dioxin | 2,3,2'-Trihydroxydiphenyl ether[d] | 434 (63), 419(11), 331 (77), 73 (100) | 16.988 | 100 | + | + |

| | Dibenzothiophene | Dibenzothiophene-sulfone[d] | 200 (7), 184 (100), 171 (7), 139 (18), 73 (4) | 17.718 | Traces | + | + |
| | | Dibenzothiophene-sulfoxide[d] | 216 (5), 200 (5), 184 (100), 147 (72), 73 (9) | 17.735 | Traces | + | + |

580  [a] Products were identified by GC-MS analysis (TMS derivatisation). Fragment ions are expressed as *m/z* values.

581  [b] When multiple oxidation products were detected their relative abundance is indicated in %.

582  [c] The ability and inability to transform each compound are shown by "+" and "-".

583  [d] Same mass spectrum as that of relevant PAH oxidation products generated by CARDO and DFDO, and previously identified based on $^1$H and $^{13}$C

584  NMR analyses (31).

585  [e] Same retention time and mass spectrum as *cis*-9,10-dihydroxy-9,10-dihydrophenanthrene produced by Pdo1 (22).

586  [f] Same retention time and mass spectrum as *cis*-1,2-dihydroxy-12-dihydroanthracene produced by Phn1 (7).

587  [g] Monohydroxylated products are most probably formed by spontaneous dehydration of the corresponding diols.

588

Table 3: Comparison of the FlnA1A2 dioxygenase activity towards fluorene and other polycyclic substrates

| Substrate | Properties of the products formed[a] | | |
| --- | --- | --- | --- |
| | retention time (min) | m/z | concentration ($\mu M$)[b] |
| Fluorene | 15.55 | 280 | 97.5 |
| Phenanthrene | 15.92 | 278 | 96.3 |
| Anthracene | 14.66 | 278 | 9.1 |
| | 16.60 | 278 | 27.3 |
| Naphthalene | 12.11 | 228 | 1.92 |
| Dibenzofuran | 13.52 | 418 | 10.1 |
| Dibenzo-p-dioxin | 14.12 | 434 | 10.0 |

[a] Characteristics of the NBB derivatives, except for dibenzufuran and dibenzo-*p*-dioxin which were analyzed as the trimehyl silyl derivatives. Anthracene yielded two isomers, one of which was identified as anthrancene 1,2-dihydrodiol (retention time: 16.60 min).

[b] Concentrations calculated in the bacterial suspension after 23 h of incubation at 25°C. Values are means of duplicates experiments. The standard error was less than 10%.